\newcommand{\bcen}{\begin{center}}
\newcommand{\ecen}{\end{center}}
\newcommand{\bfig}{\begin{figure}}
\newcommand{\efig}{\end{figure}}

\newcommand{\half}{\frac{1}{2}}
\newcommand{\bphi}{b_{\varphi}}
\newcommand{\cphi}{c_{\varphi}}

\newcommand{\brho}{b_{\varrho}}
\newcommand{\crho}{c_{\varrho}}
\newcommand{\ec}{\epsilon_c}
\newcommand{\eb}{\epsilon_b}

\newcommand{\msun}{{\rm\,M_\odot}}

\def\la{\mathrel{\hbox{\rlap{\hbox{\lower4pt\hbox{$\sim$}}}\hbox{$<$}}}}
\def\ga{\mathrel{\hbox{\rlap{\hbox{\lower4pt\hbox{$\sim$}}}\hbox{$>$}}}}

\input{epsf}
\documentstyle[11pt,aaspp4,epsfig]{article}

\slugcomment{Accepted for publication in \apj}

\lefthead{Sambhus and Sridhar}
\righthead{}

\begin{document}
\renewcommand{\theenumi}{\Alph{enumi}}
\title{Stellar orbits in triaxial clusters around black holes
in galactic nuclei}

\author{Niranjan Sambhus \\ nbs@iucaa.ernet.in}

\and

\author{S.~Sridhar \\ sridhar@iucaa.ernet.in}

\affil{Inter-University Centre for Astronomy and Astrophysics\\
Ganeshkhind, Pune 411 007, INDIA}

\begin{abstract}

We investigate the orbital structure of a model triaxial star cluster,  
centered around a supermassive black hole (BH), appropriate to galactic 
nuclei. Sridhar and Touma~(1999) proved that the presence of the BH enforces 
some regularity in the dynamics within the radius of influence of the BH.
We employ their averaging method to reduce the degrees of freedom from three 
to two. Numerical orbit integrations, together with Poincar\'e surfaces of 
section allow us to draw a global portrait of the orbital structure;  
in our calculations we employ a model cluster potential that is triaxial and  
harmonic. The averaged dynamics of the axisymmetric case is integrable,
and we present a detailed comparison of orbits in  
oblate and prolate axisymmetric potentials. Both cases 
support resonant orbits with fixed values of eccentricity, inclination, and
periapse, whose lines of nodes rotates steadily. 
We then systematically explore significantly triaxial potentials, possessing 
small oblateness, or prolateness. Resonant orbits and their families are studied
both numerically, and through secular perturbation theory. Chaos appears to be 
suppressed for all the cases we studied, and we obtain effective third integrals. 
Some of the orbits appear to reinforce the shape of the potential; we provide
phase space, as well as real space portraits of these orbits. A particularly
promising resonant orbit exists in highly prolate, triaxial  potentials.
\end{abstract} 

\newpage

\section{Introduction}

Evidence for the presence of central black holes (BH) of mass,
$10^6 - 10^{9.5}\msun$, appears strong for about a dozen galaxies 
(c.f. Kormendy and Richstone~1995). The masses  of these objects are
consistent with the mass in BHs needed to produce the observed
energy in light from quasars (Soltan~1982, Chokshi and Turner~1992).
Recent studies suggest that the dynamical influence of the BH extends 
to well outside the nuclear regions of its host galaxy, if the orbits of 
stars carry them close to the center (c.f. Merritt~1998 for a review).
In particular, dynamical phenomena in the central regions are strongly
influenced by the BH. Stars move in the combined
gravitational potentials of the BH, and the surrounding mass of stars and 
gas (for brevity, these will be referred to as the ``cluster''). 
The radius of influence of the BH ($r_h$) may be defined as radius of the
sphere within which the mass of the surrounding stars and gas is less that
of the BH.  Within $r_h$, the potential of the BH dominates the dynamics,
and the potential of the cluster may be considered to be a small perturbation.
Sridhar and Touma~(1999; hereafter referred to as ST99) studied 
stellar dynamics within $r_h$, and arrived at the following conclusions: 

\noindent
(i) Stellar orbits may be thought of as Keplerian ellipses that deform and
precess over times that are longer than orbital times by a factor $(M/M_c)$,
where $M_c$ is the mass of the cluster enclosed by the orbit. Well within $r_h$,
$M/M_c \gg 1$. 

\noindent
(ii) The slow dynamics of precessional motions may 
be understood by {\em averaging} over the Keplerian orbital periods of stars, 
which are the shortest time scales in the problem. 

\noindent
(iii) Averaging gives rise to a secular integral of motion,
$I \simeq \sqrt{GMa}$, where $a$ is the semi--major axis of the orbit. Thus the 
BH enforces some degree of regularity in the structure of orbits that lie 
within $r_h$.  

\noindent
(iv) A stellar orbit  in a time--independent cluster conserves two integrals of 
motion. These are the energy, which is an exact invariant, 
and the semi--major axis, which is secularly conserved. 

\noindent
(v) Thus clusters which may be modelled as razor--thin disks, or as
three dimensional, axisymmetric objects, give rise to averaged dynamics
that is integrable.

\noindent
(vi) Non--axisymmetric, razor--thin disks, with power law surface density
profiles have averaged dynamics, whose qualitative features are independent 
of the steepness of the power law. This occurs because, within $r_h$, the 
Kepler potential of the BH is far steeper than any self--gravitational force
due to the cluster. 

\noindent
Here we take up the thread, and consider orbits in three dimensional clusters,
which is a far more complicated problem than the cases studied
by ST99. We specifically wish to focus on the effects of three dimensionality, 
and triaxiality. Recalling from item (vi) above that the qualitative features
of orbital structure (arrangement of lens and loop orbits) are 
insensitive to the steepness of the cusp, we wish to choose a cusp 
steepness that would yield the simplest expression for the averaged 
Hamiltonian for our three dimensional problem. For the razor--thin disks
studied by ST99, the averaged Hamiltonian turned  out to be a hypergeometric
function of the eccentricity, for a general value of the cusp steepness; on
the other hand, the harmonic case yielded the simplest expressions, composed
only of elementary functions.  Therefore, in our present venture into three
dimensions, we  restrict attention to a family of triaxial, harmonic
potentials, average over the fast orbital motion, to obtain an expression for
the Hamiltonian governing slow dynamics. As equations~(4) and (5) show, 
only elementary functions occur in this Hamiltonian, which governs nearly
Keplerian, three dimensional dynamics. The phase space of the averaged 
dynamics is four dimensional, hence the dynamics is well suited to 
exploration through Poincar\'e surfaces of section. Because of all these
favourable features we consider the Hamiltonian of equations~(4) and (5)
as the ``canonical'' Hamiltonian governing three dimensional dynamics within
the radius of influence. However, these nice features may be invalid 
when the orbital radius becomes comparable to $r_h$; in particular, the 
discrepancy between true dynamics, and averaged dynamics may be greater in 
three dimensions.

In \S~2, we provide a quick introduction to action--angle variables for the
Kepler problem, and  averaging over the fast orbital frequency.  
We then study the orbital structure as a function of the energy (after
subtracting out  the Keplerian contribution). Averaging over the Keplerian
motion, promotes the  semi--major axis as a conserved quantity. For
scale--free potentials, such as  the harmonic potential, the orbital structure
is similar at all radii for which the averaging procedure is valid; this
motivates the scalings adopted  later. Orbits that reinforces the shape of the
generating potential, demands special attention. Such orbital families play a
crucial role in the  construction of self--consistent stellar dynamical models
(e.g.  Schwarzschild~1979, de Zeeuw~1996, Merritt~1999). Thus we also
study the real space structure of orbits, when they appear worthy of mention.
Averaged dynamics for oblate and prolate  axisymmetric 
cases is  discussed in \S~3. We numerically integrate the equations of motion, 
and take Poincare sections. In \S~4, we discuss the global orbital structure 
for the triaxial case, when departures from spherical symmetry are small (small
oblateness or prolateness, but quite large triaxiality). The results are
interpreted through secular perturbation theory. 
 
The reader who is interested in getting a quick glimpse of the real--space 
structure of those orbits that reinforce triaxiality, as well as 
oblateness/prolateness of the potential, is advised to consult the list of 
relevant figures and comments provided at the end of the discussion in \S~5.

\section{Formulation of orbit--averaged dynamics}

\bfig
\centerline{\hbox{\epsfxsize=6.5in\epsfbox{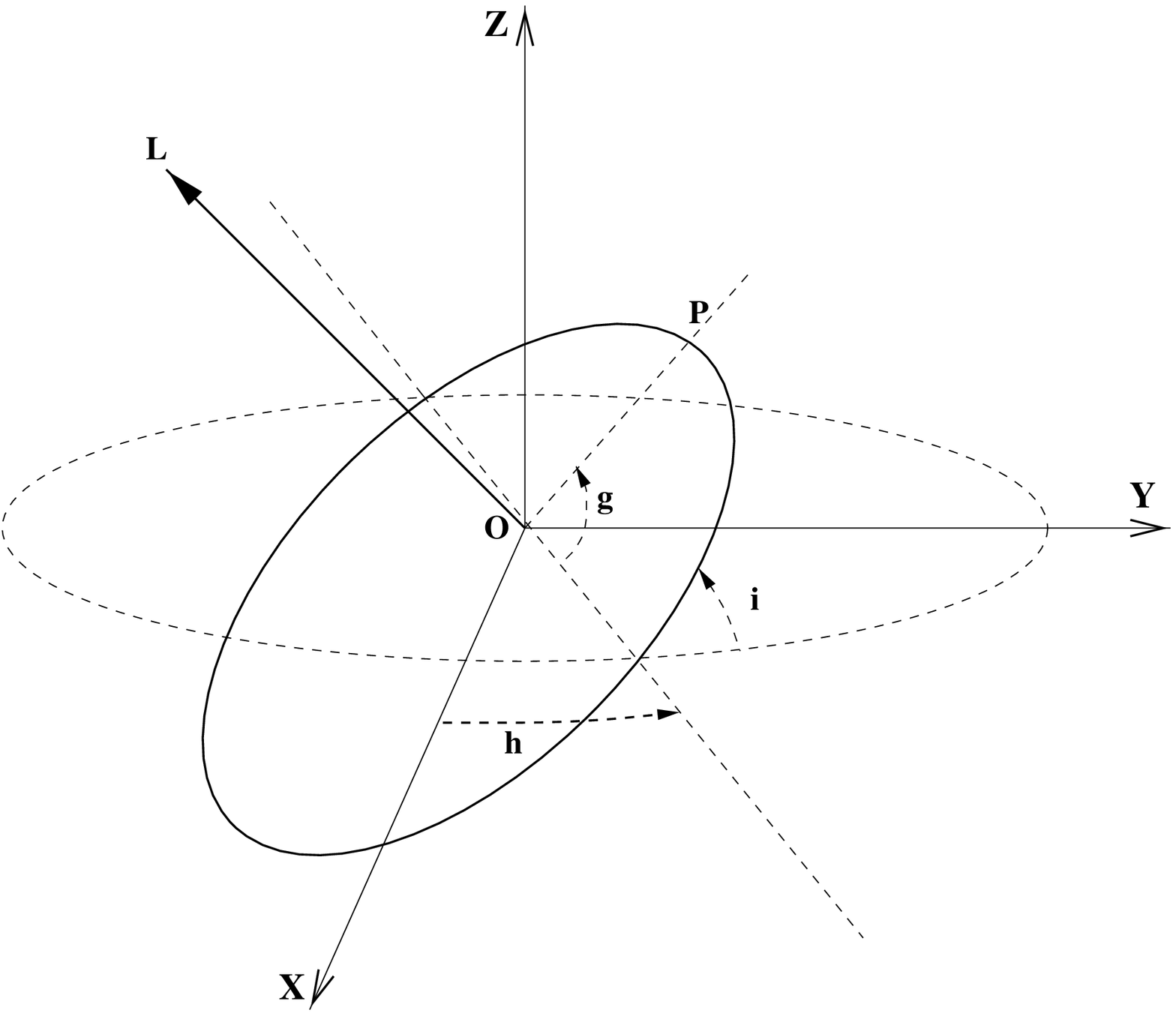}}}
\caption[Figure 1]{Orbital elements of the Kepler problem (Delaunay 
variables): the orbit is represented by the ellipse, drawn in bold
line. {\bf P} marks the pericenter. The figure shows only the four variables 
of precessional dynamics. The other two refer to the semi--major axis of the 
ellipse, and orbital phase.}
\label{fig1}
\efig
Let us locate the BH, of mass $M$,  at the origin (this is a valid assumption, 
so along as the center of mass of the surrounding cluster does not vary with
time).  We introduce action--angle variables, appropriate to the Kepler
problem,  that allow us to conveniently carry out orbit--averaging. These are
the Delaunay variables (c.f. Plummer~1960, Goldstein~1980). $I=\sqrt{GMa}$, 
$L$, and $L_z$ are the actions, where $L$ and $L_z$ are the  magnitude and 
$z$--component, respectively, of the angular momentum. We denote 
the conjugate angles by $w$, $g$ and $h$, respectively. $w$ is the orbital 
phase (``mean anomaly'') which varies from $0$ at pericenter to $2\pi$ after 
one orbit.
$g$ is the angle to the pericenter, measured from the ascending node, and 
$h$ is the longitude of the ascending node (see Figure~1). For
the Kepler problem, the only Delaunay variable that is time dependent is $w$:
$w(t) = w_0 + (G^2M^2/I^3)\, t\,$.

Let $\Phi(x, y, z)$ be the perturbing gravitational potential of the cluster. 
To average $\Phi$ over the orbital phase, it is convenient to express the spatial 
coordinates 
in terms of the Delaunay variables (c.f. Goldstein~1985):
\begin{equation}
{\left( \begin{array}{c} x \\ \\ y \\ \\ z \end{array} \right) } 
 = 
{\left( \begin{array}{ccc} 
   C_{g} C_{h} - C_{i} S_{h} S_{g} & \quad -S_{g} C_{h} - C_{i} S_{h} C_{g} & 
   \quad S_{i} S_{h}  \\ \\
   C_{g} S_{h} + C_{i} C_{h} S_{g} & \quad -S_{g} S_{h} + C_{i} C_{h} C_{g} & 
   \quad -S_{i} C_{h} \\ \\
   S_{i} S_{g}               & \quad S_{i} C_{g}            & \quad C_{i}
	\end{array} \right)} {\left( \begin{array}{c} 
   a(C_{\eta} - e) \\  \\ a\sqrt{1 - e^2}\, S_{\eta} \\  \\ 0 \end{array} 
   \right)}
\label{xyz2del}
\end{equation} 
\noindent 
where $S$ and $C$ are shorthand for sine and cosine of the angles given as 
subscript. $i$ is the angle of inclination determined by $\cos i = L_z/L$, 
and $e=\sqrt{1-L^2/I^2}$ is the eccentricity; as before, $a$ is the semi--major 
axis. $\eta$ is the eccentric anomaly, related to the orbital phase through 
$w=\eta - e\sin\eta\,$.
Having substituted the expressions for $(x, y, z)$, given in 
equation~(\ref{xyz2del}), in $\Phi(x, y, z)$, averaging is conveniently 
carried out by integrating over $\eta$:
\begin{equation}
\overline{\Phi}=\frac{1}{2\pi}\oint\,d\eta(1-e\cos\eta)\,\Phi\,.
\label{phiave}
\end{equation} 
\noindent This orbit--averaged potential, ${\overline \Phi}$,
plays the role of the Hamiltonian for slow, precessional dynamics (see ST99).
Superimposed on the slow time variation of averaged dynamics are fast 
oscillations (with frequency of order $\sqrt{GM/a^3}$), whose fractional 
amplitudes are of order $\varepsilon\sim a\,(\Phi -\overline{\Phi})/GM\,$.
Henceforth we assume that $\varepsilon\ll 1$, and ignore these fast 
oscillations. Note that, for fixed $a$, $\varepsilon$ is larger for more
eccentric orbits,  because these sample a larger spatial variation of
$\Phi\,$.  Because $\overline{\Phi}$ is independent of $w$, the conjugate
variable, $I$,  is secularly (i.e. on the average) conserved.

For investigation of orbital structure, we consider a family of triaxial, 
perturbing potentials of the form 
\begin{equation}
\Phi(x, y, z)=\frac{\Phi_0}{r_0^2}\left( x^2 + \frac{y^2}{\bphi^2}
+ \frac{z^2}{\cphi^2}\right)\,,
\label{pot}
\end{equation}
Physically, this gravitational potential may be assumed 
to arise from a homogeneous, triaxial cluster whose center coincides with 
the location of the BH. The shape parameters of the potential (viz. $\bphi$
and $\cphi$) are related to those of the density (say, $\brho$ and $\crho$)
and can be found in the literature (c.f. Chandrasekhar~1969, Binney and
Tremaine~1987).  The case when the potential is axisymmetric may be studied by
setting $\bphi=1$; oblateness, or prolateness is achieved by choosing $\cphi$
to be less than, or greater than unity, respectively. More realistic,
inhomogeneous mass distributions could be considered, but at the expense of
considerably greater algebraic complexity. 

Let $\mu=(\Phi_0 a^2/Ir_0^2)$ be a measure of the precession
frequency for orbits with semi--major axis $a$. As in ST99 , it proves to be 
convenient to define a dimensionless time $\tau=\mu t$, Hamiltonian 
$H=(\overline{\Phi}/\mu I)$, and angular momenta $\ell=L/I$ and 
$\ell_z =L_z/I$. Note that $0\leq \ell \leq 1\,$, and $-1\leq \ell_z 
\leq 1\,$. Averaging the potential of equation~(\ref{pot}) over 
orbital phase, we obtain the dimensionless Hamiltonian  
\begin{equation}
H(\ell, \ell_z, g, h) = \half (5 - 3\ell^2) \;+\; \ec H_{c}(\ell, 
\ell_z, g)  \;+\;  \eb H_{b}(\ell, \ell_z, g, h)\,, 
\label{htri}
\end{equation}
\noindent where 
\begin{eqnarray}
H_c(\ell, \ell_z, g) & = & \frac{1}{4}\left(1-\frac{\ell_z^2}{\ell^2}\right)
\left( 5 - 3 \ell^2 - 
5 (1 - \ell^2)\, C_{2g}\, \right)\,,\nonumber \\[1em]
H_b(\ell, \ell_z, g, h) & = & \left(\frac{5}{2} - 2\ell^2\right) 
\left(C_g S_h + \frac{\ell_z}{\ell} C_h S_g \right)^2 + \frac{\ell^2}{2} 
\left(- S_g S_h + \frac{\ell_z}{\ell} C_h C_g \right)^2\,, 
\label{hbc}
\end{eqnarray}
\noindent and $\eb= (\bphi^{-2} -1)$, $\ec= (\cphi^{-2} -1)$. This is the 
model two degree--of--freedom Hamiltonian that will be studied in the rest of 
this paper. The equations of motion are, 
 \begin{equation}
\dot{\ell}=-\frac{\partial H}{\partial g}\,,\quad
\dot{g}=\frac{\partial H}{\partial \ell}\,;\quad\quad\quad
\dot{\ell_z}=-\frac{\partial H}{\partial h}\,,\quad
\dot{h}=\frac{\partial H}{\partial \ell_z}\,,
\label{eom}
\end{equation}
\noindent where the over--dots refer to $(d/d\tau)\,$. 
We integrate these equations numerically using a fourth--order, adaptive 
step--size, Runge--Kutta scheme, and take surfaces of section at constant 
values of either of the angles $g$, or $h$. Thus we follow the 
the deformations of Keplerian ellipses of fixed semi--major axes, over time
scales much longer than the orbital periods. This technique will, in principle,
reveal all of the phase space structure of the orbits. We will also supplement
the numerical calculations with analytical estimates, wherever necessary. 
Note that such an assault on orbits in triaxial potentials would have been
impossible, without the presence of the extra integral of motion that
averaging promotes.  

The Hamiltonian, given in equations~(\ref{htri}) and (\ref{hbc}),
is a constant of motion. The orbital structure shows strong dependence 
on the value of this quantity, and henceforth we will refer to it as the 
energy ($E$). Since, for the averaged dynamics, Kepler energy is constant,
this refers refers to the average value of $\Phi$ over the orbit in question.
 The simplest case is that of spherical symmetry, when $\epsilon_b=\epsilon_c=0$.
Then $H=(5-3\ell^2)/2$, so that $\dot{\ell}=\dot{\ell_z}=\dot{h}=0\,$, and 
$\dot{g}=-3\ell\,$: the Keplerian ellipses precess rigidly, at a constant rate
in a retrograde sense, maintaining an invariant orbital plane.  

\section{Orbits in the axisymmetric case}
 
When $\bphi=1$,  the potential of equation~(\ref{pot}) is 
axisymmetric about the $z$--axis. Hence $L_z$ is conserved even for the 
unaveraged dynamics. For slow dynamics, this implies that $\ell_z$ is an 
integral of motion, 
and we are left with a one  degree--of--freedom system, whose dynamics is
obviously integrable. Setting $\eb=0$, from 
equations~(\ref{htri}) and (\ref{hbc}), we obtain the following expression 
for the axisymmetric Hamiltonian, 
\begin{equation}
H(\ell, g; \ell_z) = \half (5 - 3\ell^2) \;+\;  
\frac{\ec}{4}\left(1-\frac{\ell_z^2}{\ell^2}\right)
\left( 5 - 3 \ell^2 - 
5 (1 - \ell^2)\, C_{2g}\, \right)\,.
\label{haxi}
\end{equation}
\noindent where $\ell_z$ is now regarded as a constant parameter.
Note that $\ec$ is positive for oblate configurations, and negative for the 
prolate ones. The nodal precession rate given by
\begin{equation}
\dot{h} = \frac{\partial H}{\partial \ell_z}= -\half\ec\frac{\ell_z}{\ell^2}
\left[10\,S^2_g\, (1-\ell^2) + 2\ell^2\right]\,, 
\label{hdotaxi}
\end{equation}
\noindent shows that the sign of $\dot{h}$ is opposite to that of the product,
 $\ec\ell_z$; hence nodal precession is prograde (retrograde)
for prolate (oblate) potentials.

To obtain a global picture of the dynamics, it is instructive to plot an 
$\ell - g$ (Poincar\'e) surface of section taken at some fixed value of $h$; 
Figures~2a and 2b show two such sections for oblate ($\cphi < 1$) and prolate   
($\cphi > 1$) potentials, respectively. 
\bfig[htt]
\centerline{\mbox{\epsfig{figure=fig2a.ps,width=3.0in,angle=270}}}
\vskip 0.1in
{Fig. 2a.\,\, $\ell - g$ surface--of--section for oblate axisymmetric potential.
 $\cphi = 0.8$ and energy $E= 2.48$ }
\label{fig2a}
\vskip 0.2in
\centerline{\mbox{\epsfig{figure=fig2b.ps,width=3.0in,angle=270}}}
\vskip 0.1in
{Fig. 2b.\,\, $\ell - g$ surface--of--section for prolate axisymmetric potential. 
$\cphi = 2.5$ and energy $E= 0.5$ }
\label{fig2b}
\efig

\bfig[htt]
\centerline{\mbox{\epsfig{figure=fig2c.ps,width=3.0in,angle=270}}}
\vskip 0.1in
{Fig. 2c.\,\, $\ell - g$ surface--of--section for oblate axisymmetric potential.
 $\cphi = 0.7$ and energy $E= 2.2$ }
\label{fig2c}
\vskip 0.2in
\centerline{\mbox{\epsfig{figure=fig2d.ps,width=3.0in,angle=270}}}
\vskip 0.1in
{Fig. 2d.\,\, $\ell - g$ surface--of--section for prolate axisymmetric potential.
$\cphi = 2.5$ and energy $E= 0.6$ }
\label{fig2d}
\efig

From these figures, it may be seen that the dynamics is
organised around the elliptic fixed points, located at $g=\pi/2\mbox{ and } 3\pi/2$.
These are special orbits for which $\ell$ and $g$ are constant in time, 
whereas $h$ precesses steadily. Using the Hamiltonian of equation~(\ref{haxi})
in equations~(\ref{eom}), it is straightforward to verify that these 
fixed point orbits (FPOs) are described by, 
\begin{equation}
\ell^4=\frac{5\epsilon_c\ell_z^2}{3+4\epsilon_c}\,,\quad\quad\quad
g=\pi/2\;\;\mbox{or}\;\;3\pi/2\,,\quad\quad\quad
\dot{h}=4\epsilon_c\ell_z -5\epsilon_c\sqrt{\frac{3+4\epsilon_c}
{5\epsilon_c}}\,. 
\label{axifp}
\end{equation}
\noindent Before we proceed to examine the dynamics in more detail, let us 
pause to consider some implications of equations~(\ref{axifp}) for 
FPOs (it is useful to recall that $0\leq\ell\leq 1\,$, 
$-1\leq\ell_z\leq 1$, and $\cos i = \ell_z/\ell\,$).

\noindent 1. One end of the FPO sequence is set by $\ell_z = 0$, which is the 
$z$ axial orbit (unless $3 + 4\epsilon_c = 0$, which is the case $\cphi = 2$).

\noindent 2. Because $\ell^2\propto \cos^2i$, the orbits become rounder 
as their inclination decreases (i.e. as the orbital plane approaches the
equator). 

\noindent 3. The other end of the FPO sequence is reached when either $i=0$, 
or $\ell = 1$, whichever occurs first. This depends on whether 
$5\epsilon_c/(3+4\epsilon_c)$ is less than, or greater than unity
(equivalently, when $\cphi$ is greater than, or less than $1/2\,$).
There is an additional requirement that $5\epsilon_c/(3+4\epsilon_c)\geq 0$, 
which implies that no FPOs of the kind allowed by equation~(\ref{axifp})
exist, for prolate potentials which have $1<\cphi<2\,$.

\noindent 4. {\it Oblate: $0<\cphi<1/2$}: $\ell = 1$ is reached before 
$i = 0$, so there are no equatorial FPOs.

\noindent 5. {\it Oblate: $1/2<\cphi<1$}: $i=0$ is reached before $\ell = 1\,$.
When $\cphi =1/2$, equation~(\ref{axifp}) implies that $ell^2 = \cos^2 i$ 
for the FPO; thus both $i=0$ and $\ell=1$ are reached simultaneously.

\noindent 6. {\it Prolate: $\cphi > 2$}: $\ell = 1$ is reached before 
$i = 0$, so there are no equatorial FPOs.

The Poincar\'e section may also be regarded from a different perspective. 
For the completely integrable dynamics of the axisymmetric problem, one 
may invert equation~(\ref{haxi}), to express $\ell_z$ as a function of 
$\ell$, $g$ and $H$,
\begin{equation}
\ell_z = \pm\,\ell\left( 1 - \frac{1}{\ec}\,\frac{ 6\ell^2 + 4H - 10}
{\ell^2(5C_{2g} - 3) + 5(1- C_{2g})}\right)^{1/2}\,.
\label{lzaxi}
\end{equation}
\noindent We may now regard $H$ as a constant parameter, with the $\ell - g$
dynamics controlled by the Hamiltonian equal to $(-\ell_z)$, and ``time''
measured by $h$. Figures~2a and 2b may be reproduced by plotting   
the isocontours of $\ell_z$ in the $\ell - g$ plane. It is straightforward to 
verify the stability of the orbits of equation~(\ref{axifp}) by Taylor expanding 
$\ell_z$ about the fixed point values of $\ell$ and $g$. 
The FPOs parent a family of 
orbits for which both $\ell$ and $g$ librate periodically, and $h$ precesses
non--uniformly in time. In real space, these orbits
fill (``hollowed--out'') conical regions whose axes coincide with the $z$--axis.
The  phase flows in the $\ell - g$ plane are structurally similar to 
those of a pendulum. Hence there are also other orbits for which $g$ circulates, and
these are distinguished from the former by a separatrix. Figures~2c and 2d show
two representative sections of such orbits.

\bfig[htt]
\centerline{\mbox{\epsfig{figure=fig2e.ps,width=3.0in,angle=270}}}
\vskip 0.1in
{Fig. 2e.\,\,  $\ell - g$ surface--of--section for prolate axisymmetric potential.
$\cphi = 1.5$ and energy $E= 2.3$ }
\label{fig2e}
\efig

For prolate potentials
with $1 < \cphi < 2$, the dynamics is quite different, as is revealed by 
the surface of section, shown in Figure~2e; $g$ circulates for all orbits. 
This difference in behaviour is connected with the stability of the $z$--axis
orbit. Let us test the $z$--axis orbit for stability to eccentricity variations
which, however, maintain $\ell_z=0$. We Taylor expand the right side of 
equation~(\ref{haxi}) about $\ell=0$, and $g=\pi/2$. Setting $g=\pi/2 + \delta
g$, we obtain
\begin{equation}
\frac{5}{2}(1+\ec) - H = (\frac{3}{2} + 2\ec)\ell^2 + \frac{5}{2}\ec\,
(\delta g)^2\,.
\label{haxiexp}
\end{equation}
\noindent The energy of the $z$--axis orbit is $H = 5(1 + \ec)/2$, which is 
the maximum allowed energy in the oblate case. Therefore, the left side is 
always non negative, as are the coefficients of $\ell^2$ and $(\delta g)^2$; thus
the $z$--axis orbit is always stable. For the prolate case, the energy of the 
$z$--axis orbit, $H = 5(1 + \ec)/2$, is the minimum allowed energy. Hence the 
left side is always non positive. The coefficient of $(\delta g)^2$ is negative,
so the issue of stability depends on the sign of the coefficient of $\ell^2$.
It is clear that the $z$--axis orbit is stable for $\ec<-3/4$, and unstable
when $-3/4<\ec<0$.

\section{Orbits in the triaxial case.}

The shape of the potential is given by two axes ratios, $\bphi$ and $\cphi$.
The definition of the Delaunay variables gives a privileged status to the 
$z$--axis, so we let this be the axis of symmetry for the study 
of orbits in the axisymmetric case. In particular, we set $\bphi=1$, and 
chose $\cphi < 1$ and $\cphi > 1$ for oblate and prolate cases respectively. 
Whereas this is the most convenient choice for the dynamics, it implies that
the ordering of the lengths of  the $x$, $y$ and $z$ axes are different in the 
oblate and prolate cases. This feature carries over to triaxial configurations 
as well. Hence we allow $\bphi$ and $\cphi$ to take any positive value, and
define the ``triaxiality'' to be  $T = (a_1^2 - a_2^2)/(a_1^2 - a_3^2)$, where
 $(a_1, a_2, a_3)$ are equal to the set $(1, \bphi, \cphi)$, after the latter
has been rearranged (if necessary) in descending order of
magnitudes. With this convention, for  axisymmetric configurations, $T=0$ for the
oblate case, and $T=1$ for the  prolate case.

The equations of motion (eqn.~\ref{eom}), resulting from the Hamiltonian 
of equations~(\ref{htri}) and (\ref{hbc}), are 
\begin{eqnarray}
\dot{\ell} & = & \ec \left[\frac{5}{2}\left(1 -\frac{\ell_z^2}{\ell^2}\right) 
(1 - \ell^2) S_{2g} \right] \,\,\, - 5 \eb \left[(1 - \ell^2) (C_g S_h +
\frac{\ell_z}{\ell} C_h S_g) (- S_g S_h + \frac{\ell_z}{\ell} C_h C_g) 
\right]\,,\nonumber \\[1em]
\dot{\ell_z} & = & - \eb\left[(5 - 4\ell^2) (C_g S_h + \frac{\ell_z}{\ell}
C_h S_g) (C_g C_h - \frac{\ell_z}{\ell} S_h S_g)\right]\nonumber\\  
& & +\eb\left[\ell^2 (-S_g S_h +
\frac{\ell_z}{\ell} C_h C_g) (S_g C_h + \frac{\ell_z}{\ell} S_h C_g) \right]
\,,\nonumber \\[1em]
\dot{g} & = & - 3\ell + \frac{1}{2}\ec\left[\frac{\ell_z^2}{\ell^3}(5 - 3\ell^2 
- 5(1 - \ell^2) C_{2g}) + \ell (1 - \frac{\ell_z^2}{\ell^2}) (5 C_{2g} -3)\right]
\nonumber\\
& & - \eb\left[(C_g S_h + \frac{\ell_z}{\ell} C_h S_g) (4\ell C_g S_h +
5\frac{\ell_z}{\ell^2} C_h S_g) + \ell S_g S_h (-S_g S_h + \frac{\ell_z}{\ell} 
C_h C_g)\right]
\,,\nonumber \\[1em]
\dot{h} & = & - \frac{1}{2}\ec \frac{\ell_z}{\ell^2}(5 - 3 \ell^2 -5 (1 -
\ell^2) C_{2g}) \nonumber \\
& & + \eb\left[(5 - 4\ell^2)\frac{1}{\ell} C_h S_g (C_g S_h +
\frac{\ell_z}{\ell} C_h S_g)  +  \ell C_h C_g (-S_g S_h + \frac{\ell_z}{\ell} 
C_h C_g)\right]\,.
\label{eomtri}
\end{eqnarray}
\noindent We integrate these numerically, and take 
Poincar\'e sections 
by strobing in either of the angles $g$, or $h$. 
In this paper, we restrict our study to triaxial configurations that are 
close to spherical; thus $\bphi$ and $\cphi$ are close to unity (equivalently, 
$\eb$ and $\ec$ are close to zero). This allows us to use
perturbation theory to understand the results of our numerical computations. 

As in the axisymmetric case we study the orbital structure as a function of 
the (constant value of) Hamiltonian, which we refer to as the ``energy''.
Recall that the Hamiltonian is equal to the (scaled) potential of the cluster. 
The scaling simply means that we have chosen to explore the dynamics of 
nearly Keplerian orbits with unit semi--major axis. Hence the lowest energy
orbits are circular, and the highest energy orbits very eccentric.
Below we explore both oblate and prolate triaxial cases. As representative 
cases, we consider $(\bphi, \cphi)$ equal to $(0.99, 0.96)$
and $(0.99, 1.04)$, for the oblate and prolate triaxial potential. It may be 
verified that the triaxiality is quite large, namely $T\simeq 0.254$, 
and $0.804$, respectively.

\subsection{Low energy orbits; $\ell\la O(1)$}

Recall that $\ell_z$ was a constant of motion in the axisymmetric case;
a Poincar\'e section, strobed in $g$ would have given horizontal rows of dots
in the $\ell_z - h$ plane. Hence this surface of section is very useful in 
revealing the effects of triaxiality, as may be seen in  
Figures~3a and 3b, which show a pendulum-like dynamics of 
the $\ell_z - h$ motions. The figures do not, of course, reveal that 
the periapse angle, $g$, always circulates (but this has been verified by 
plotting $\ell - g$ sections, and by following the time evolution of 
$g$). A glance at the equations of motion~(\ref{eomtri}) shows that $|\dot{g}|
\sim O(1)$, whereas $|\dot{h}|\sim O(\ec)\,\ll |\dot{g}|$. Therefore, 
it is useful to imagine the motion of the 
Keplerian ellipses on two well--seperated timescales. At nearly constant values 
of $h$ and $i$ (i.e. on a nearly fixed orbital plane), the precessing ellipse 
fills a circular annulus. On a longer timescale, the orbital plane itself 
either circulates or librates in $h$,  while oscillating in inclination. As
in the  axisymmetric case, oblate triaxial configurations give rise to
retrograde  precession of  $h$ while prograde precession results in the
prolate case. In Figure~3a, the librating and circulating orbits are 
long--axis and short--axis tubes, respectively; in Figure~3b, 
the roles are reversed. These orbits are also found in more general
triaxial galactic potentials with constant density cores. Their 
persistence in the case when a black hole is present may 
be understood qualitatively by noting that these tube orbits are round
($\ell\la O(1)$), so they are not destabilized by the black hole 
(note that the lower $\ell$ orbits of \S~4.2 are completely different).

The dynamics can be understood by averaging over the fast angle
$g$.  Equations~(\ref{htri}) and ~(\ref{hbc}), when averaged over 
$g$ give
\begin{equation}
\left< H\right>_g = \half (5 - 3\ell^2) \left[ 1 \; + \;
\frac{\ec}{2}\left(1-\frac{\ell_z^2}{\ell^2}\right) \; +\;
\frac{\eb}{2}\left(S_h^2 \,+\, \frac{\ell_z^2}{\ell^2}\,C_h^2
\right)\,\right]\,,
\label{htgav}
\end{equation} 
\noindent
for the Hamiltonian describing the slower dynamics of $\ell_z$ and $h$; it is 
straightforward to verify that the isocontours of this Hamiltonian reproduce 
the surfaces of sections shown in Figures~3a and 3b. 
In this approximation, $\ell$ emerges as a secular invariant (i.e. no secular 
variation in eccentricity), and the $g$--averaged dynamics reduces to a 
one degree of freedom system in the variables, $\ell_z$ and  $h$.\footnote{A
more systematic approach  to averaging over the fast angle uses first order
perturbation theory, whose  primary advantage over equation~(\ref{htgav}) is
that the $O(\epsilon_c)$  oscillations in $\ell$ are well represented.}
\bfig[htt]
\centerline{\mbox{\epsfig{figure=fig3a.ps,width=3.0in,angle=270}}}
\vskip 0.1in
{Fig. 3a.\,\, $\ell_z - h$ surface--of--section (strobed at $g=\pi/2$) for 
oblate triaxial potential. $\bphi = 0.99$, $\cphi = 0.96$ and energy $E= 1.8$
 } \label{fig3a}
\vskip 0.2in
\centerline{\mbox{\epsfig{figure=fig3b.ps,width=3.0in,angle=270}}}
\vskip 0.1in
{Fig. 3b.\,\, $\ell_z - h$ surface--of--section (strobed at $g=\pi/2$) for
 prolate triaxial potential. $\bphi = 0.99$, $\cphi = 1.04$ and energy $E=
1.8$ } \label{fig3b}    
\efig
The orbits corresponding to circulations of $h$ (short and 
long axes tubes, for oblate and prolate cases, respectively) have larger
$|\ell_z|$,  so are more nearly {\em equatorial}. The orbit fills a 
region of space that is  toroidal and non axisymmetric. The approximate axis of 
symmetry is along the $z$--axis, and the torus itself could be thought of as  
somewhat pinched along the intermediate axis ($y$ for oblate and $x$ for prolate).

The orbits for which $h$ librates (and $\ell_z$ flips sign), have smaller 
values of $|\ell_z|$, so these are more nearly {\em polar} (long and 
short axes tubes, for oblate and prolate cases, respectively). Let us first
consider the parent polar orbit with $\ell_z=0$. For the oblate/prolate cases,
the orbit  lies in a plane perpendicular to the long/short axis. Within the
plane, $g$  circulates rapidly and nearly uniformly, while $\ell$ shows
equally rapid,  small amplitude oscillations; hence the orbit fills an
elliptical annular region.  Figure 4 shows projections of a  polar orbit
on the three principal planes for a prolate case. The orbit lies essentially
in the plane of the long and the intermediate axes, so reinforces the shape of
the potential to some extent.

\bfig[htt]
\centerline{\mbox{\epsfig{figure=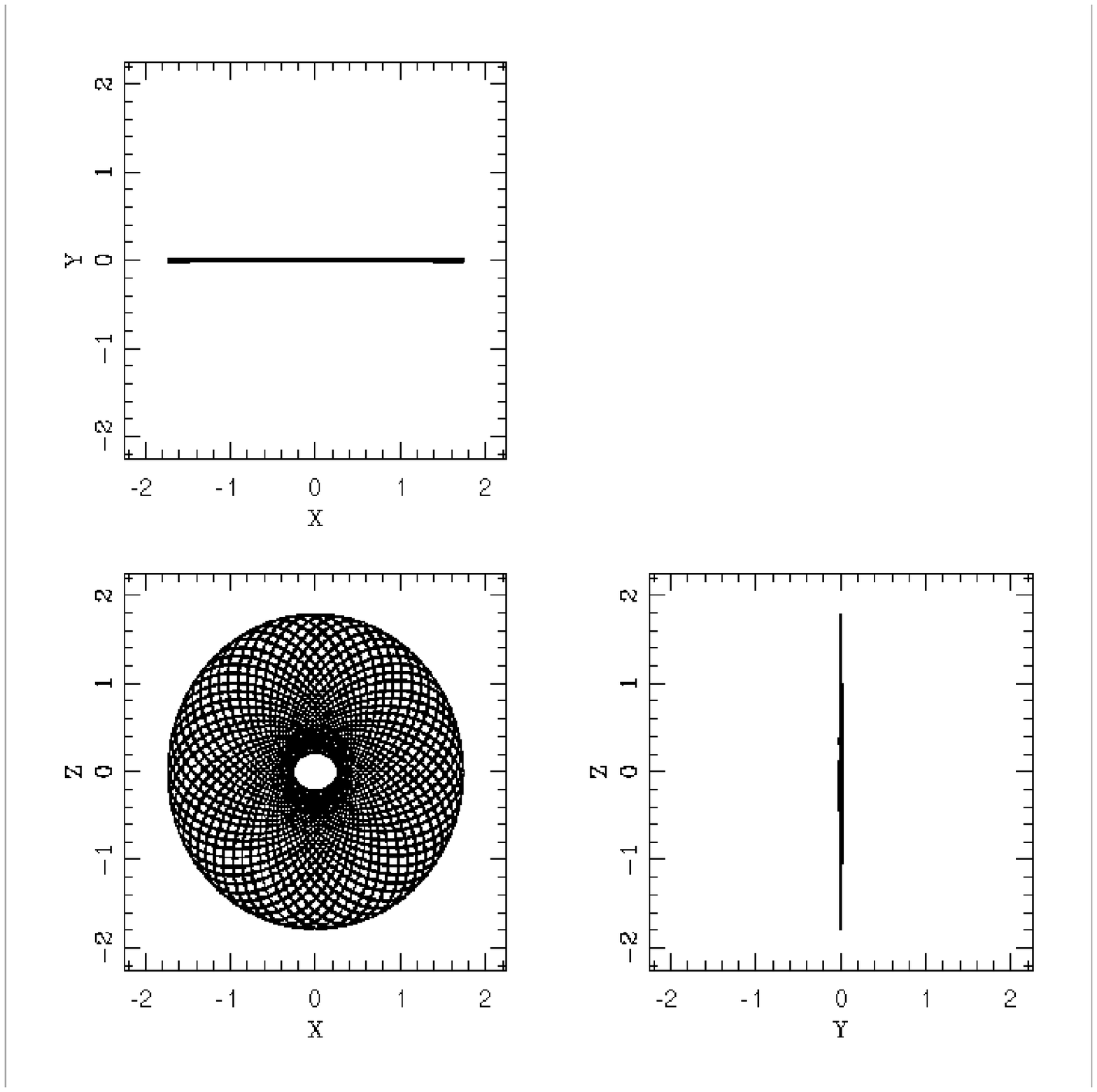,width=6.5in,angle=0}}}
\vskip 0.1in
{Fig. 4.\,\, Real space projections of a polar orbit in prolate triaxial
 potential. $\bphi = 0.99$, $\cphi = 1.04$ and energy $E = 1.8$ }
\label{fig4}    
\efig

\subsection{High energy orbits: $\ell\sim O(\sqrt{\ec})$}

High energy orbits are characterised by low angular momentum values, hence are 
all quite eccentric. In general, $\ell\sim O(\sqrt{|\ec|})$. This regime of 
energy admits a certain scaling, that allows us to parametrise the dynamics by 
the single quantity; $B=\eb/|\ec|$. Let us introduce the scaled
variables,
\begin{equation}
l=\frac{\ell}{\sqrt{|\ec|}}\,,\quad\quad
l_z=\frac{\ell_z}{\sqrt{|\ec|}}\,,\quad\quad
K=\frac{1}{|\ec|}\left(H-\frac{5}{2}\right)\,,\quad\quad
s=\sqrt{|\ec|}\,\tau\,,
\label{scale}
\end{equation}
\noindent where $(l, l_z)$ are conjugate to $(g, h)$, and $K$ and $s$ are
the new Hamiltonian and time, respectively, for this very slow dynamics of
highly eccentric motions. Substituting the new variables in
equations~(\ref{htri}) and (\ref{hbc}), we obtain
\begin{equation}
K=-\frac{3}{2}l^2 \;+\; sgn(\ec) \frac{5}{4}\left(1-\frac{l_z^2}{l^2}\right)
\left(1-C_{2g}\right) \;+\; \frac{5B}{2}\left(C_g S_h + \frac{l_z}{l}
S_g C_h\right)^2 \; +\; O(\epsilon_b,\epsilon_c)\,.
\label{hscale}
\end{equation}
\noindent where $sgn$($\ec$) is $\pm$, for $\ec$ positive/negative.
Whereas the scaling is useful in restricting dependence to
one parameter, $B$, the dynamics still concerns two coupled degrees of freedom, 
and no further reduction is, in general, possible. However, $K$ is 
algebraically simpler to handle than $H$ and we use it to understand the
results of  numerical integrations of the full equations of motion
(eqns.~\ref{eomtri}).  The dynamics of these highly eccentric orbits differs
in the oblate and prolate  triaxial potentials. 

\subsubsection{Oblate triaxial potential}
\bfig[htt]
\centerline{\mbox{\epsfig{figure=fig5a.ps,width=3.0in,angle=270}}}
\vskip 0.1in
{Fig. 5a.\,\,  $\ell_z - h$ surface--of--section (strobed at $g = \pi/2$)
for oblate triaxial potential. $\bphi = 0.99$, $\cphi = 0.96$ and energy
$E= 2.48$ }
\label{fig5a}
\vskip 0.2in
\centerline{\mbox{\epsfig{figure=fig5b.ps,width=3.0in,angle=270}}}
\vskip 0.1in
{Fig. 5b.\,\,  $\ell - g$ surface--of--section (strobed at $h = \pi/2$) for 
oblate triaxial potential. $\bphi = 0.99$, $\cphi = 0.96$ and energy 
$E= 2.48$ }
\label{fig5b}    
\efig
Figures~5a and 5b are Poincar\'e sections, strobed in $g$ and $h$ respectively;
it is important to note that the $\ell_z - h$ section is taken at 
$g = \pi/2\;(\mbox{mod} 2\pi)$, but the precise value of the strobing in $h$, 
for the $\ell - g$ section is immaterial. Figure~5a is very similar to
Figure~3a, with the notable addition of elliptic fixed points at $h = 0$ and
$\pi$, located at non zero values of $|\ell_z|$. Moreover, Figure~5b shows
them to be triaxial versions of the axisymmetric fixed points, discussed in
\S~3 (see Figure~2a). Recall that, in the axisymmetric case, the fixed point
orbit corresponded to the steady rotation, about the $z$--axis, of an
inclined, rigid Keplerian ellipse. \bfig[htt]
\centerline{\mbox{\epsfig{figure=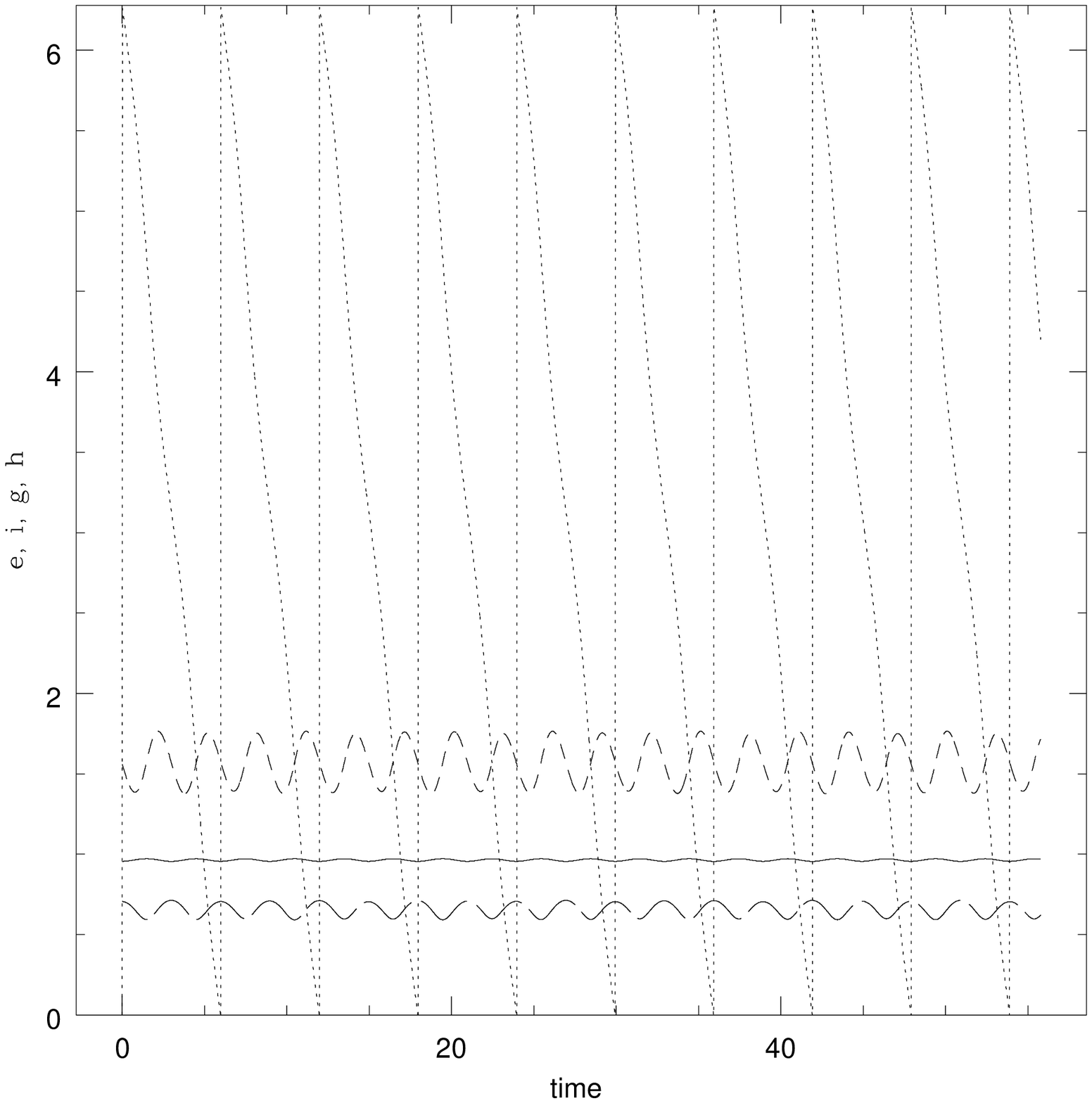,width=4.5in,angle=0}}} \vskip 0.1in
{Fig. 6.\,\, Temporal behaviour of $i, e, g, h$ for the $\ell - g$ fixed point
orbit in oblate triaxial potential with $\bphi = 0.99, \cphi = 0.96, \mbox {
and } E = 2.48$. The long--dashed, solid, short--dashed, and the dotted
curves, correspond to $i, e, g, h$, repsectively. } \label{fig6} \efig The
main effect of triaxiality is to introduce oscillations into quantities that
were steady in the axisymmetric case. A question of interest is whether these
oscillations change the shape of the orbit in a manner that reinforces the
shape of the potential. Figure~6 shows the behaviour in time of $g$, $h$, $i$
and $e$,  where it may be seen that the librations of $g$ (about $\pi/2$, or
$3\pi/2$) occur at {\em twice} the frequency at which $h$ precesses. 
Physically, this can be thought of as an $m=2$ perturbation of an 
axisymmetric problem. Below we provide an analysis of this perturbation, 
and derive expressions the quantitative effects of triaxiality on 
various quantities of interest, such as the energy at which the FPO appears, 
the amplitudes of oscillation induced in $\ell$, $\ell_z$, and $g$.

Let us rewrite $K$ 
(eqn.~\ref{hscale}) as the sum of two terms, $K_1$ and $K_2$; $K_1$
contains terms that are independent of $h$, and $K_2$ consists of the remaining
$h$ dependent terms.
\begin{eqnarray}
K &=&  K_1 + K_2\,,  \nonumber\\[1em]
K_1 &=& -\frac{3}{2}l^2 \;+\;\frac{5}{4}\left(1-\frac{l_z^2}{l^2}
\right)\left(1-C_{2g}\right)\;+\;\frac{5B}{4}\left(C_g^2 + \frac{l_z^2}{l^2}
S_g^2\right)\,\,, \nonumber\\[1em]
K_2 &=& \frac{5B}{4}\left(-C_g^2 C_{2h} + \frac{l_z^2}{l^2}S_g^2 C_{2h} + 
\frac{l_z}{l} S_{2g} S_{2h} \right)\,\,\,.
\label{kaxitri}
\end{eqnarray}
\noindent $K_1$ may be viewed as an effective axisymmetric system
 that reduces to the actual axisymmetric case as $B \rightarrow 0$. Clearly,
 $l_z$ is an integral of motion for this system, which we henceforth denote
as $l_{z_0}$. The Hamiltonian $K_1$ admits elliptic fixed points at   
\begin{equation} g_0 = \frac{\pi}{2}\,, \frac{3\pi}{2}\,\,, \qquad\quad 
l_0 = \left[\frac{5}{3}\left(1 - \frac{B}{2}\right) l_{z_0}^2 \right]^{1/4}\,,
\label{lgfixk1}
\end{equation} 
\noindent for which the nodal angle $h$ circulates uniformly in the retrograde 
sense at a frequency, $\dot{h_0} = -\sqrt{15(1 - B/2)}$, which is independent 
of the value of $l_{z_0}$. When $B$ equals zero, this expression
coincides with the $\dot{h_0}$  of eqn~(\ref{axifp}), upto $O(\ec)$.
The fixed point value of the energy, $K_{1_0}$ is fixed by the value of the 
parameter $l_{z_0}$. The minimum value of energy at which such a fixed point is 
admitted is then obtained by sustituting the above fixed point values of 
$l$ and $g$ in the Hamiltonian $K_1$, and also noting that the maximum value
$l_z$ can  take is equal to $l$. Thus,
\begin{equation}
K_{1_{min}} = \frac{5}{2} ( B - 1)\,,
\label{oeminscale}
\end{equation}
\noindent from which we obtain (using equation~(\ref{scale})),
\begin{equation}
E_0 = \frac{5}{2}  +  |\ec| K_{1_{min}}\,\,=\,\,\frac{5}{2} (1 - \ec + \eb) .
\label{oemin}
\end{equation}
The Hamiltonian $K_2$ acts as a perturbation on this effective axisymmetric 
system. Below we present  a first order  analysis of the perturbed fixed point 
orbit. To this end, let $(l_1, g_1, l_{z_1}, h_1)$ be the perturbations to the 
fixed point values of the effective axisymmetric system,   
$(l_0, g_0, l_{z_0}, h_0)$. Recalling that $(l_0, g_0, l_{z_0})$ are constant
in time, and $h_0 = -\sqrt{15(1 - B/2)}s$, we note that 
the observed oscillations in the values of the dynamical variables 
(see Figure~6) will be described by ($l_1, l_{z_1}, g_1, h_1$). 
Hamilton's equations of motion, given by, 
\begin{eqnarray}
\frac{d l_1}{d s} &=& -\, l_1
\left(\frac{\partial^2 K_1}{\partial l\partial g}\right)_0   -\, 
 l_{z_1} \left(\frac{\partial^2 K_1}{\partial l_z\partial g}\right)_0   -\, 
g_1 \left(\frac{\partial^2 K_1}{{\partial g}^2}\right)_0   -\,
\left(\frac{\partial K_2}{\partial g}\right)_0\,, \nonumber\\[1em] \frac{d
g_1}{d s} &=&  l_1 \left(\frac{\partial^2 K_1}{{\partial l}^2}\right)_0  +\,
l_{z_1} \left(\frac{\partial^2 K_1}{\partial l_z\partial l}\right)_0  +\,  g_1
\left(\frac{\partial^2 K_1}{\partial g\partial l}\right)_0  +\,  
\left(\frac{\partial K_2}{\partial l}\right)_0\,, \nonumber\\[1em] \frac{d
l_{z_1}}{d s} &=&  -\,  \left(\frac{\partial K_2}{\partial h}\right)_0\,,
\nonumber\\[1em] \frac{d h_1}{d s} &=& l_1
\left(\frac{\partial^2 K_1}{\partial l\partial l_z}\right)_0  +\, 
l_{z_1} \left(\frac{\partial^2 K_1}{{\partial l_z}^2}\right)_0  +\,  g_1
\left(\frac{\partial^2 K_1}{\partial g\partial l_z}\right)_0  +\,
\left(\frac{\partial K_2}{\partial l_z}\right)_0\,,
\label{obtrieom}
\end{eqnarray}
\noindent admit the following solutions,
\begin{eqnarray}
l_1 &=& \frac{1}{4} {\left(\frac{5}{3}\right)}^{3/4}B
{\left(1 - \frac{B}{2}\right)}^{-1/4} l_{z_0}^{-1/2} \cos 2\omega_h s \,,
\nonumber\\ g_1 &=& \frac{1}{2} {\left(\frac{5}{3}\right)}^{1/4}B{\left(1 -
\frac{B}{2}\right)}^{-3/4} l_{z_0}^{-1/2} \sin 2\omega_h s \,, \nonumber\\
l_{z_1} &=& \frac{B}{4} {\left(1 - \frac{B}{2}\right)}^{-1} \cos 2\omega_h s
\,, \nonumber\\ h_1 &=& -\frac{B}{4} {\left(1 - \frac{B}{2}\right)}^{-1/2}
\left[ \half + \sqrt{\frac{5}{3}} l_{z_0}^{-1} \right] \sin 2\omega_h s \,, 
\nonumber\\ h_0 &=& \omega_h s\,,  \label{ooscillations} \end{eqnarray}
\noindent where, $\omega_h = -\sqrt{15(1 - B/2)}$. Thus it is clear that the 
frequency of the small amplitude oscillations  
is twice that of the mean rotation of $h$. Moreover, $l$ and $l_z$ oscillate 
{\em in phase}, whereas the oscillations of $g$ about its fixed point value 
and those of $h$ about its mean rotation are {\em in phase}. Note that the 
amplitude of these oscillations decreases to zero
as $B \rightarrow 0$, as expected for the axisymmetric case. 
Table~1 compares the predictions of our first--order theory with the results 
of numerical integrations of the exact equations of motion, for two different 
values of $B$. The  smaller the value of $B$, the better is the match 
between the observed and predicted values. 

Figure~7 shows the projections of one such FPO along the three 
symmetry planes defined by, the long and the intermediate axes (X--Y plane), 
the intermediate and the short axes (Y--Z plane), and the long and the short
axes  (X--Z plane). The orbit can be viewed as a highly elongated ellipse, 
rotating about the minor axis (here Z--axis) in a nonuniform manner. The orbit 
makes a nonzero inclination that exhibits a small libration about its mean value.
 Corresponding to these librations, is a deformation of the ellipse 
itself, resulting in  periodic changes of its eccentricity. It is apparent from 
the figures that the orbit fills a conical region in space. The orbital extent 
is maximum along the long axis, and minimum along the short axis. 

The orbit corresponding to the separatrix of Figure~5b, is shown in projection
in Figure~8. This is seen to fill a rounder region of space than the fixed 
point orbit. It is interesting to note that both orbits (shown in Figure~7 and 
Figure~8) have shapes that appear to largely reinforce the triaxiality, as well 
as the  oblateness of the potential.  
\begin{table}[ht]
\caption{The observed and the predicted values of $E_0$, and the relative 
amplitudes of the fluctuations in $\ell$, $\ell_z$ and $g$  for the fixed point 
orbit in oblate triaxial potential. Note that smaller the value of $B$, better 
is the match in values.} 
\vskip 0.2in
\begin{tabular}{ccccccc}
\hline
  & $B$ & $T$ & $E_0$ & $(l_1 / l_0)_{max}$ & $(l_{z_1} / l_{z_0})_{max}$ & 
$(g_1 / g_0)_{max}$ \\
\hline
\hline
  &  &  &  &  &  &  \\
  & $0.118$ & $0.127$ &   &   &   &   \\
  &  &  &  &  &  &  \\
Observed &    &    & $2.334$ & $0.548$ & $0.031$ & $0.059$ \\
  &  &  &  &  &  &  \\
Predicted &    &    & $2.312$ & $0.541$ & $0.031$ & $0.052$ \\
  &  &  &  &  &  &  \\
\hline
  &  &  &  &  &  &  \\
  & $0.239$ & $0.254$ &   &   &   &   \\
  &  &  &  &  &  &  \\
Observed &    &    & $2.356$ & $0.170$ & $0.065$ & $0.120$ \\
  &  &  &  &  &  &  \\
Predicted &    &    & $2.331$ & $0.110$ & $0.067$ & $0.110$ \\
  &  &  &  &  &  &  \\
\hline
  &  &  &  &  &  &  \\
\end{tabular}
\end{table}

\bfig[htt]
\centerline{\mbox{\epsfig{figure=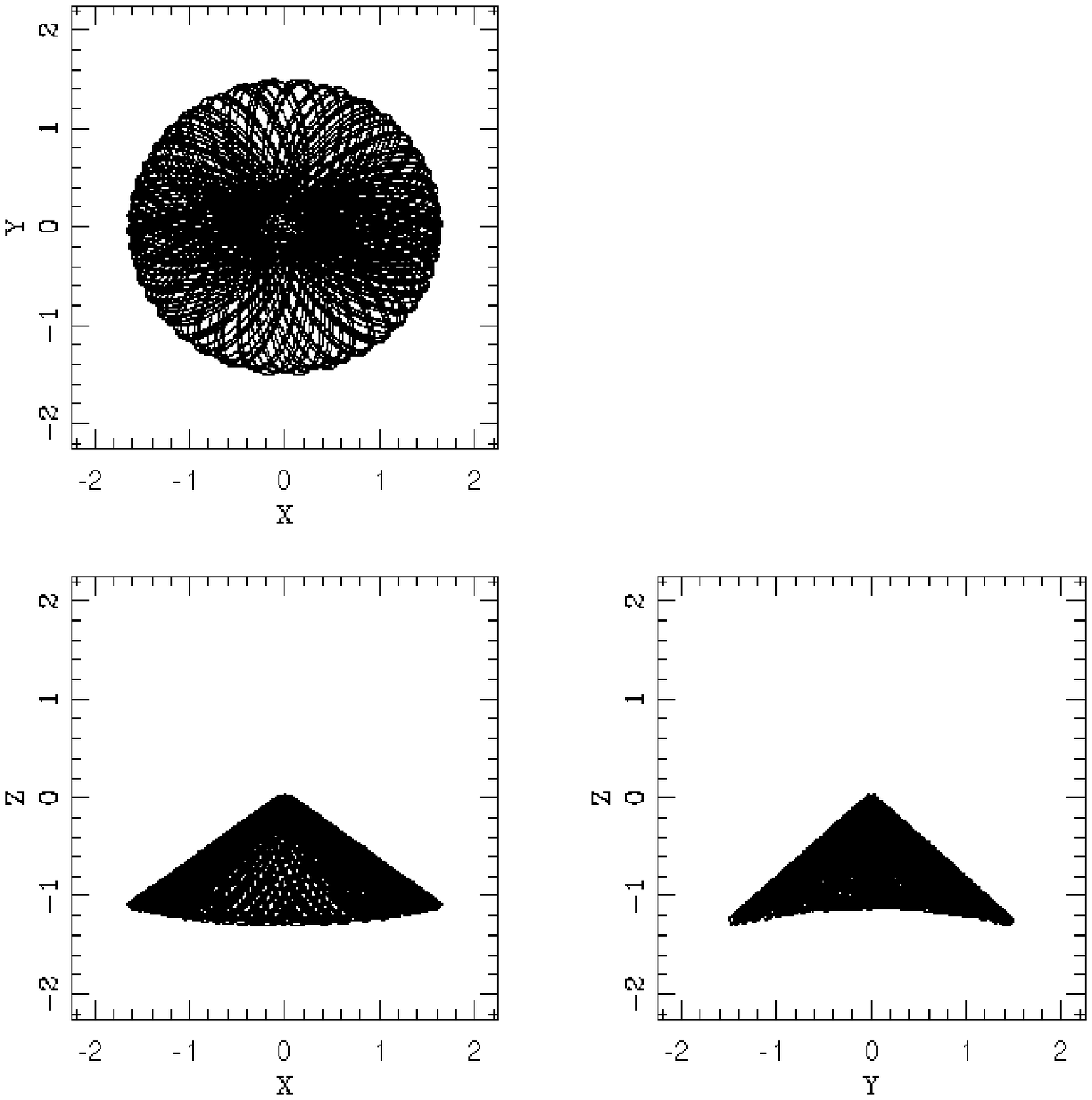,width=6.5in,angle=0}}}
\vskip 0.1in
{Fig. 7.\,\,  Real space projections of the $\ell - g$ FPO 
in oblate triaxial potential. $\bphi = 0.99$, $\cphi = 0.96$ and energy $E =
2.48 $  }
\label{fig7}    
\efig
\bfig[htt]
\centerline{\mbox{\epsfig{figure=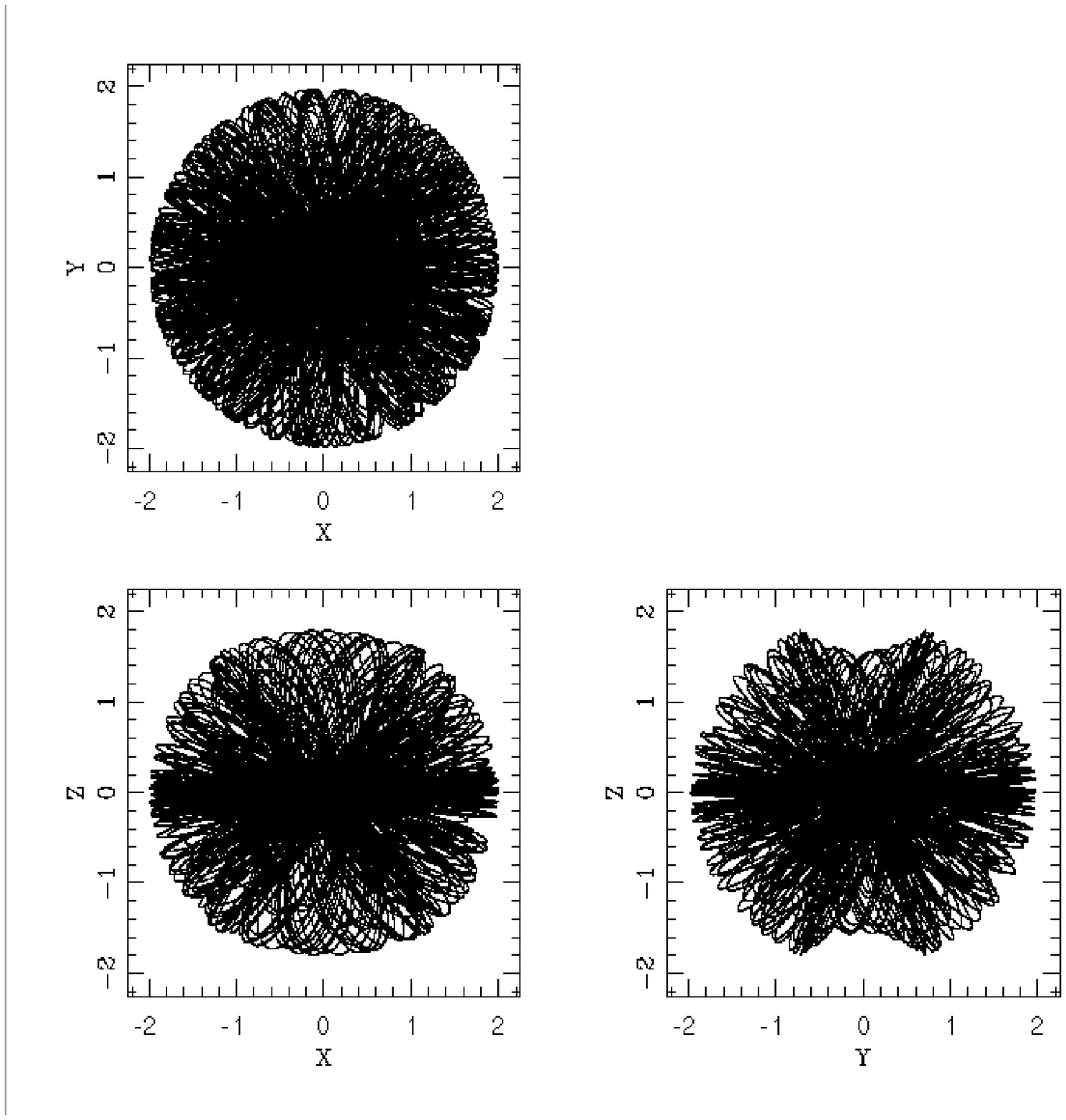,width=6.5in,angle=0}}}
\vskip 0.1in
{Fig. 8.\,\,  Real space projections of the separatrix orbit in the oblate 
triaxial case. $\bphi = 0.99$, $\cphi = 0.96$ and energy $E = 2.48$}
\label{fig8}    
\efig

\subsubsection{Prolate triaxial potential}

\bfig[htt]
\centerline{\mbox{\epsfig{figure=fig9a.ps,width=3.0in,angle=270}}}
\vskip 0.1in
{Fig. 9a.\,\, $\ell - g$ surface--of--section (strobed at $h = 0$) for
 prolate triaxial potential. $\bphi = 0.99$, $\cphi = 1.04$ and energy $E=
2.3$  } \label{fig9a}
\vskip 0.2in
\centerline{\mbox{\epsfig{figure=fig9b.ps,width=3.0in,angle=270}}}
\vskip 0.1in
{Fig. 9b.\,\, $\ell_z - h$ surface--of--section (strobed at $g = \pi/2$)
for prolate triaxial potential. $\bphi = 0.99$, $\cphi = 1.04$ and energy $E= 2.3$ }
\label{fig9b}    
\efig
Figures~9a and 9b show Poincar\'e sections, strobed at $h = 0$ and $g= \pi/2$
respectively for a triaxial potential, whose prolateness is small ($\cphi$
larger than, but close to unity).
Both  sections show FPOs surrounded by islands of stability.
Recall that prolate axisymmetric potentials, with $1<\cphi< 2$ do not admit
elliptic fixed points, for any value of the energy (compare Figure~9a with
Figure~2e). Hence this fixed point is a purely triaxial phenomenon,
which is conveniently described in the $\ell_z - h$ surface of section.
Now compare Figure~9b with Figure~3b. Both the configurations have
the same triaxiality, but different energies. Note that the stable fixed point at
$\ell_z=0$ in Figure~3b has changed its character to an unstable fixed point, as seen
in Figure~9b. The latter figure also shows emergence of two stable fixed points,
which were absent at low energies.
It should be emphasised that these fixed points appear only above a certain
minimum energy, $E_0$. The occurance of these fixed points can be explained by
adhereing to an analogy with parametric resonance. 

Consider the Hamiltonian $K$, given in equation~(\ref{hscale})\footnote{ This 
choice of the Hamiltonian already implies that one is looking at potentials
with small  $|\ec|$}. Expanding $K$ about  the fixed point $(l_z, h) = (0,0)$,
and keeping  terms upto quadratic order in $l_z$ and $h$, we write
\begin{eqnarray}
K &=& K_g + K_h ,\\  \nonumber
K_g &=& -\left(\frac{3}{2}l^2 \; +\;\frac{5}{4} (1-C_{2g})\right),\\  \nonumber
K_h &=&  \left(\frac{5(1 + B)}{4 l^2}(1 - C_{2g})\right) l_z^2\; + \; 
\left(\frac{5 B (1 + C_{2g})}{4}\right) h^2\; + \; \left(\frac{5 B S_{2g}}{2l}
\right) l_z h\,.
\label{hsplit}
\end{eqnarray}
\noindent $K_g$ and $K_h$ may be viewed as the two coupled oscillators.
$K_h$ is quadratic in the small quantities, $l_z$ and $h$, hence$|K_h|\ll
|K_g|$, and we may make the approximation, 
\begin{equation}
\dot{l}\simeq -\frac{\partial K_g}{\partial g}\,,\qquad\qquad
\dot{g}\simeq \frac{\partial K_g}{\partial l}\,.
\label{lgpend}
\end{equation}
\noindent These equations are identical to those of a pendulum, and are 
readily solved. Of the two possible kinds of motion, libration and circulation, 
it is the latter that turns out to be relevant for our problem. We now 
regard $l$ and $g$ as given functions of time in $K_h$, and solve the 
following equations, 
\begin{equation}
\dot{l_z}= -\frac{\partial K_h}{\partial h}\,,\qquad\qquad
\dot{h}= \frac{\partial K_h}{\partial l_z}\,.
\label{lzhpend}
\end{equation}
\noindent to obtain $l_z(t)$ and $h(t)$. Thus the dynamics has been reduced 
to a master--slave system, where the $\ell_z - h$ oscillations are driven by
the $\ell - g$ motion. As the energy of the system is increased, the forcing
frquency of the $K_g$ pendulum changes, leading to a parametric instability in
the $\ell_z - h$ system. This renders the $\ell_z - h$ fixed point
hyperbolic, simultaneously giving birth to two new stable fixed points, as
shown in Figure~9b. The  separation between the new, stable, fixed points
increases with energy. Below, we estimate $E_0$ by considerations appropriate to
the onset of this parametric instability. We note that $K_g$
is nearly constant, and eliminate $l$ in favour of $K_g$ and $g$. This is 
substituted in the expression for $K_h$ in equation~(\ref{hsplit}), to obtain 
\begin{equation}
K_h = \left( -\frac{15}{2}\frac{(1 - C_{2g})(1 + B)}{(4K_g + 5(1 - C_{2g}))} 
\right) l_z^2 + \left(\frac{5B(1 + C_{2g})}{4}\right) h^2 + 
\left( \frac{5\sqrt{3}B S_{2g}}{\sqrt{-2(4K_g + 5(1 - C_{2g})}}\right) l_z h .
\label{lzhham}
\end{equation}
\noindent Because the angle $g$ circulates, we average the 
Hamiltonian, given in equation~(\ref{lzhham}) over $g$, to obtain,
\begin{equation}
\left<K_h\right>_g = \left( -\frac{3(1 + B)}{2}\left( 1 - 
\sqrt{\frac{2 K_g}{5 + 2 K_g}}
\right)\right) l_z^2 + \frac{5 B}{4} h^2\,.
\label{lzhosc}
\end{equation}
\noindent The cross term in $l_z-h$ has dropped out, leaving  behind a simple 
harmonic oscillator, whose frequency of small oscillations is,
\begin{equation}
\omega_h = \sqrt{-\frac{15}{2}B(1 + B)\left(1 - \sqrt{\frac{2 K_g}{5 + 
2 K_g}}\right)}\,.
\label{lzhfreq}
\end{equation}
\noindent This should be compared with the frequency of circulation of 
the $K_g$ pendulum,
\begin{equation}
\omega_g = \sqrt{-24 \left(K_g + \frac{5}{4}\right)}.
\label{l2gfreq}
\end{equation}
\noindent We expect parametric resonance when $\omega_g\simeq 2\omega_h$. 
Imposing this condition on equations~(\ref{lzhfreq}) and (\ref{l2gfreq}), 
gives us an fifth order, algebraic equation for $K_g$, whose one real root 
gives us a prediction for the minimum energy, $E_0^{pre} = (5/2) + |\ec| K_g$, 
at which the $(\ell_z=0, h=\pi)$ fixed point goes unstable, and bifurcates 
into two new elliptic fixed points. Table~2, given below, compares
$E_0^{pre}$, with the minimum energy, $E_0$, obtained from numerical
integrations of the full equations of motion. \begin{table}[ht] \caption{The
observed and the predicted values of the minimum energy $E_0$ for the fixed
point orbit for various prolate triaxial potentials. The superscripts $obs$
and $pre$ refer to the observed and the predicted values. } \vskip 0.2in
{\large \begin{tabular}{cccc} \hline  $B$  &  $T$    &   $E_0^{obs}$   &  
$E_0^{pre}$  \\ \hline
\hline
   &    &    &   \\
$0.523$  & $0.670$  & $2.36$ & $2.39$ \\
   &    &    &   \\
$0.354$  & $0.754$  & $2.31$ & $2.34$ \\
   &    &    &   \\
$0.269$  & $0.804$  & $2.27$ & $2.30$ \\
   &    &    &   \\
\hline
   &    &    &   \\         
\end{tabular}}
\end{table}

\bfig[htt]
\centerline{\mbox{\epsfig{figure=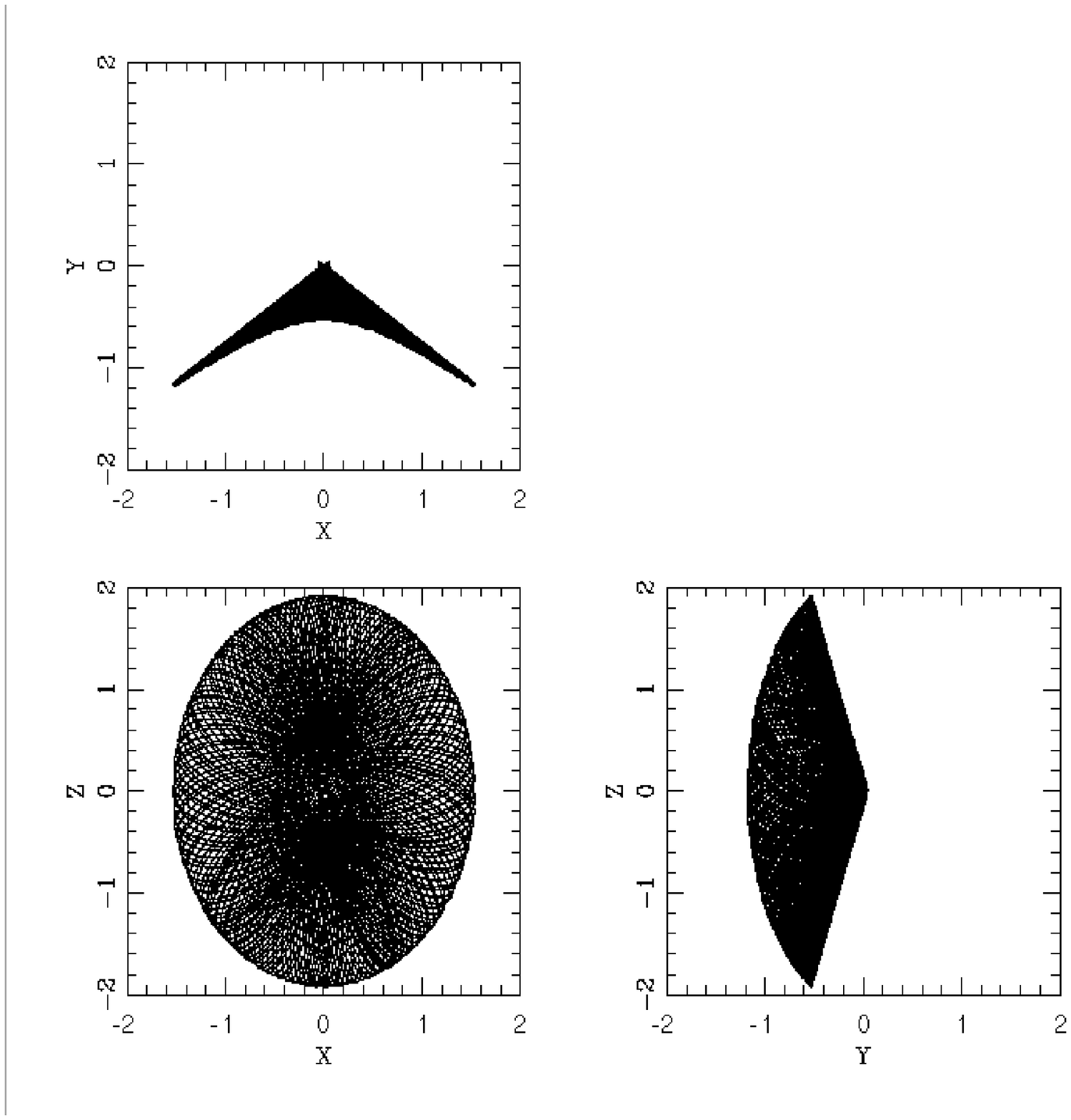,width=6.5in,angle=0}}}
\vskip 0.1in
{Fig. 10.\,\,  Real space projections of the $\ell - g$ FPO in 
prolate triaxial potential. $\bphi = 0.99$, $\cphi = 1.04$ and energy $E = 2.3$}
\label{fig10}    
\efig
Figure~10 shows  projections of one such FPO on the principal 
planes. The orbit reinforces the prolateness, as well as the triaxiality
of the potential. 

\subsubsection{Triaxial potential, with large prolateness}

\bfig[htt]
\centerline{\mbox{\epsfig{figure=fig11.ps,width=3.0in,angle=270}}}
\vskip 0.1in
{Fig. 11.\,\, $\ell - g$ surface--of--section (strobed at $h = \pi/2$) for
 prolate triaxial potential. $\bphi = 0.99$, $\cphi = 2.5$ and energy $E=
0.5$  } \label{fig11}
\efig
When $\cphi > 2$, the prolate, axisymmetric potential admits a fixed point 
orbit, which survives the introduction of some triaxiality.  
Figure~11 is a Poincar\'e section strobed at $h = \pi/2$ for a prolate triaxial 
configuration with $\cphi = 2.5$ and $\bphi = 0.99$. This figure is very 
similar to Figure~2b. The projections onto the principal planes reveals that
the  orbit strongly reinforces both the triaxiality and the prolateness of the 
potential.  
\bfig[htt]
\centerline{\mbox{\epsfig{figure=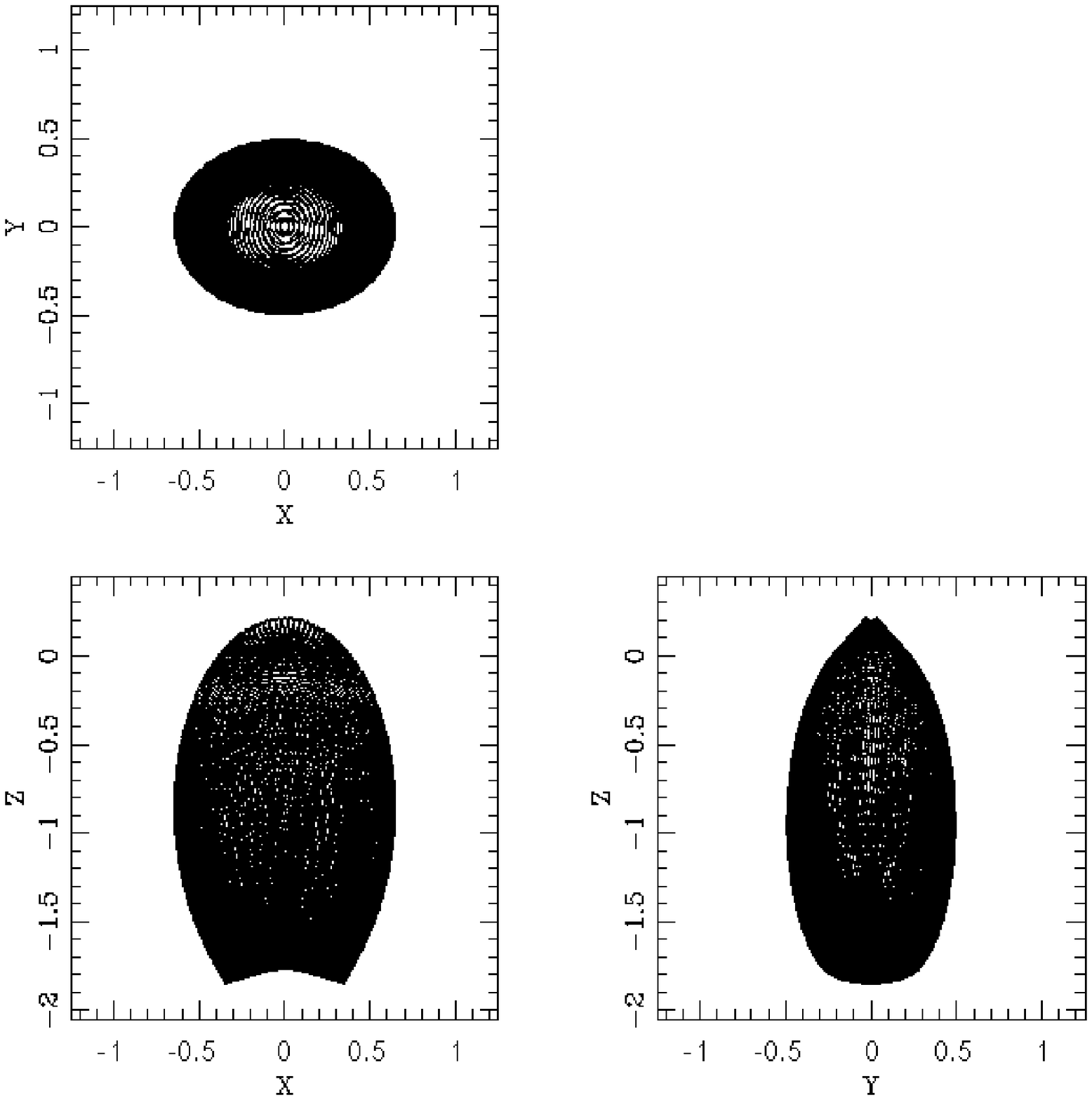,width=6.5in,angle=0}}}
\vskip 0.1in
{Fig. 12.\,\,  Real space projections of the $\ell - g$ FPO in 
prolate triaxial potential. $\bphi = 0.99$, $\cphi = 2.5$ and energy $E = 0.5$ }
\label{fig12}    
\efig
Figure~12 shows real space projections of such a FPO.

\section{Discussion}

In the nuclear regions of galaxies, stars move in the combined gravitational 
fields of the central, supermassive black hole (BH), and the self--gravity of 
the cluster of stars. When the cluster is triaxial, the only integral of 
motion that orbits respect is the energy, and  orbital structure is
notoriously difficult to understand. 
We have applied the orbit--averaging method of ST99 to study stellar orbits 
that lie within the radius of 
influence of BH, where the gravitational potential of the cluster is a 
small perturbation to the Keplerian potential of the BH.
Averaging over the orbital phase of the fast, nearly Keplerian motion 
promotes an additional integral of motion: for a time independent
perturbation, this integral is the semi--major axis. Hence the problem reduces
in dimension to the  dynamics of the two precessional degrees of freedom. For
this first effort at addressing the effect of triaxiality, we modeled the
potential of the cluster by a triaxial, harmonic  potential (the averaging
method itself applies to general potentials); the rationale for this choice
is discussed in some detail in the Introduction. We formulated the
equations of motion using  Delaunay variables (which are the action--angle 
variables of the Kepler problem), and integrated orbits numerically; 
appropriate Poincare sections provide global information on the dynamics.

The averaged dynamics of the axisymmetric case is completely integrable, 
and the orbital structure can be easily analysed. Both oblate and prolate
axisymmetric  potentials admit resonant orbits that parent families of their
own. See \S3  for a detailed listing of the properties of these orbits. 
Kozai~(1962) discussed a similar problem in the context of the motion of 
asteroids perturbed by Jupiter. More recently, Holman, Touma and
Tremaine~(1997) have also considered the axisymmetric problem in the context
of the dynamics of a planet orbiting the star, 16~Cyg~B, being perturbed by
the distant stellar companion, 16~Cyg~A.\footnote{In both cases, the perturber
was modeled as a circular ring, and the tidal field was truncated at
quadrupolar order, giving rise to an axisymmetric, harmonic potential of the
form given in our equation~(\ref{pot}), with $\Phi_0 < 0$, $\bphi=1$, and
$\cphi=1/\sqrt{2}$.} 

In the triaxial case, the averaged dynamics admits two integrals of motions;
the energy, which is exactly conserved, and the semi--major axis, which is 
secularly conserved. We undertook a systematic study of the nearly spherically 
symmetric case which, however, was quite highly triaxial. Chaos appeared to be,
suppressed, and we identified appropriate effective third integrals
in the several cases we studied. This could be a result of the reduction 
in degrees offered by averaging over the fast Keplerian motion. 
Averaging will be an increasingly invalid procedure as the size of the orbit
approaches $r_h$; it will not surprise us if the three dimensional dynamics
is significantly chaotic for these larger orbits. 

Our study of the averaged dynamics of the triaxial case has turned up 
several resonant orbits, many of which appear to reinforce the triaxiality,
as well as the oblateness/prolateness of the potential. Below we provide a
brief travel guide.

\noindent 1. {\em Polar orbits in prolate, triaxial potential} : Projections on 
principal planes given in Figure~4. Poincar\'e section in Figure~3b. 
Discussed in \S~4.1. Orbit fills elliptical annulus in the plane of the
intermediate and  long axes.

\noindent 2. {\em Resonant orbit in oblate triaxial potential} : Projections on 
principal planes given in Figure~7. Poincar\'e section in Figures~5a and 5b. 
Discussed in \S~4.2.1. Orbit fills a hollow conical region, of elliptical 
cross section, whose axis is oriented along the short axis of the potential.

\noindent 3. {\em Separatrix orbit in oblate triaxial potential} : 
Projections on principal planes given in Figure~8. 
Poincar\'e section in Figures~5b. Mentioned in \S~4.2.1. Orbit fills a three
dimensional box--like region, whose dimensions are compatible with the shape
of the potential. Orbit spends a lot of time in the equatorial plane.

\noindent 4. {\em Resonant orbit~1 in prolate triaxial potential} : 
Projections on principal planes given in Figure~10. 
Poincar\'e section in Figures~9a and 9b. Discussed in \S~4.2.2. Orbit present
only for small prolateness. Shaped like a butterfly. 

\noindent 5. {\em Resonant orbit~2 in prolate triaxial potential} : 
Projections on principal planes given in Figure~12.
 Poincar\'e section in Figures~11. Discussed in \S~4.2.3. Orbit present only
for large prolateness. Fills a hollow gourd--like region. Highly reinforces
triaxiality. A promising candidate for constructing self--consistent models. 

\section{Acknowledgements}

We thank an anonymous referee for laboring with a dense manuscript,
and patiently criticising incomplete, sometimes incorrect, statements.
NS thanks the Council of Scientific and Industrial Research, India, for
financial support through grant  2--21/95(II)/E.U.II.

\end{document}